\documentclass[showpacs,aps,prb,groupaddress,preprint]{revtex4}
\usepackage{amsmath}
\usepackage{amssymb}
\usepackage{graphicx,color}
\usepackage{epstopdf}

\begin{document}

\title{Tunable photo-galvanic effect on topological insulator surfaces via proximity interactions}

\author{Yuriy G. Semenov}
\affiliation{Department of Electrical and Computer Engineering, North Carolina State University, Raleigh, NC 27695-7911}

\author{Xiaodong Li}
\affiliation{Department of Electrical and Computer Engineering, North Carolina State University, Raleigh, NC 27695-7911}

\author{Ki Wook Kim}\email{kwk@ncsu.edu}
\affiliation{Department of Electrical and Computer Engineering, North Carolina State University, Raleigh, NC 27695-7911}

\begin{abstract}
An unusual photo-galvanic effect is predicted on the topological insulator surface when its semi-metallic electronic spectrum is modified by an adjacent ferromagnet.  The effect is correlated with light absorption in a wide frequency range (from a few to hundreds of meV) and produces a pronounced response that is not only resonant to the photon energy but also tunable by an external electrical bias.  The exceptionally strong peak photocurrent of the order of $\mu$A/cm may be achieved at elevated temperatures with the illumination power of 1 W/cm$^2$ in the THz range on Bi$_2$Se$_3$.  These advantages could enable room-temperature detection of far-infrared radiation.


\end{abstract}

\maketitle

To date the photo-galvanic effect (PGE), i.e., direct current
generation via light absorption in a structure with broken inversion
symmetry, has been used as a unique experimental tool to study
kinetic phenomena. Three different mechanisms are known to trigger
the PGE: (i) the drag of electron gas by
photons,~\cite{Ryvkin70,Zeeger}   (ii) the imbalance in the
excitation of electrons with opposite
momenta,~\cite{Ivchenko95,Diehl09}   and (iii) the shear of
electron non-equilibrium distribution in the momentum space in the
course of relaxation to
equilibrium.~\cite{Ivchenko95,Ivchenko08}   In this regard,
the spin-orbit interaction leads to a specific manifestation of the
latter two cases as it has been demonstrated previously that the
transfer of optical spin polarization can realize a direct
current.~\cite{Ivchenko95}  However, the induced current is
generally of low density due to the secondary role of spin-orbit
coupling in the formation of band structure. Other approaches of PGE
invariably suffer from the same fate of feeble response.

One potential exception may be the topological insulators (TIs) $-$
the material with spin-orbit interaction so strong that it inverses
the conduction and valence bands and exhibits numerous unusual
phenomena deviating from the conventional perception about the
magneto-electric
effects.~\cite{Zhang2009,Tse10,Zhang2010,Hasan2010,Qi2011}
It is plausible to expect the peculiar properties of TI surface
states also modify the PGE, particularly when the time-reversal
symmetry is broken by external fields. For instance, absorption of
circularly polarized light was recently shown to generate a direct
current in the surficial region of the TI.~\cite{Hosur2011}
Generally, this \textit{circular} PGE is conditioned by the
spin-momentum interlocked nature of the TI surface states. To get a
nonzero response, however, a magnetic field $B$ must break the
three-fold rotational symmetry that results in an extremely weak
photocurrent of $\sim 10$ fA/cm under the excitation power of 1
W/cm$^{2}$ and $B=10$ T. A more pronounced effect ($\sim 10$ nA/cm
at $T=15$~K) was reported in Ref.~\onlinecite{Mclver2011}, where
the TI surface isotropy is broken by the obliquely incident light
with an excitation energy much higher than the TI band gap. While
the spin-orbit interaction is the key to the circular PGE, the weak
low-symmetry interactions limit the density of dc transport to a low
level in both cases as before.

In this study we draw attention to another, much stronger mechanism
of PGE that stems from the band structure asymmetry at the interface
between a TI and a proximate {\em insulating} or {\em dielectric}
ferromagnet (FM). The basic principle of this effect is illustrated
in Fig.~1. Here, the FM is assumed to be magnetized along the $y$
axis that results in the shift of the TI surface Dirac cones along
the $x$ direction away from the $\Gamma$ point in the momentum
(i.e., $k$) space.~\cite{Hasan2010,Qi2011}    We focus,
however, on one consequence of this shift beyond the contact point
that is manifested in the contortion of the conic shapes due to the
spin-independent, quadratic-in-$k$ term [see the dashed curves in
Figs.~1(b) and 1(c)]. Specifically, the states on the right side of the
shifted Dirac
cones in the figure are farther from the $\Gamma $ point, in which the $%
k^{2} $ term in the Hamiltonian increases more rapidly for both
conduction and valence bands. For the n-type TI plotted in
Fig.~1(b), the optical excitation is suppressed for the states on
the left side of the Dirac cone, since both initial and final states
are situated below the electrochemical potential $\mu $. On the
other hand, the right side of the cone allows the transitions
between resonant states via the same photons because they (i.e., the
electronic states) are located astride $\mu $. Consequently, a
significant imbalance induced in the momentum (i.e., group velocity)
distribution of the excited electron ensemble produces a net
photocurrent. Similarly, the p-type TI [Fig.~1(c)] gives a nonzero
photocurrent along the opposite direction because the optical
excitations on the right of the Dirac cones are suppressed instead.

As a didactic but practically important example of the approach, we
start with a discussion on optical absorption in the TI/FM
heterostucture before proceeding to the PGE. The corresponding
states are described by the effective Hamiltonian of the low-energy
TI surface electrons in an exchange interaction with a proximate
magnet. Keeping the linear and quadratic terms of the
two-dimensional electron momentum $\mathbf{k} =(k_{x},k_{y},0)$, the
Hamiltonian takes the form~\cite{Shan2010}
\begin{equation}
H=\hbar v_{F}[\vec{\mathbf{\sigma }}\times \mathbf{k}]\cdot \mathbf{\hat{z}}%
+Dk^{2}+\mathbf{G}\cdot \vec{\mathbf{\sigma }},  \label{H1}
\end{equation}%
where $\mathbf{\hat{z}}=(0,0,1)$ is the unit vector normal to the TI
surface, $v_{F}$ is the Fermi velocity, $\vec{\mathbf{\sigma }}=(\sigma
_{x},\sigma _{y},\sigma _{z})$ is the vector Pauli matrices of the electron
spin, and $D$ is the material parameter for the quadratic term. The last
term in Eq.~(\ref{H1}) describes the exchange interaction with the FM in
terms of the effective exchange field $\mathbf{G}=\xi \mathbf{M}$, where $%
\mathbf{M}$ denotes the FM magnetization and $\xi $ is proportional
to the exchange integral.~\cite{Yokoyama2010,Kong2011}
Since the exchange energy involves only the electron spin operators,
its effect can be simply regarded
as a momentum renormalization for the first term in Eq.~(\ref{H1}), with $k%
\mathbf{_{y}}\rightarrow k\mathbf{_{y}}+G_{x}/\hbar v_{F}$ and $%
k_{x}\rightarrow k_{x}-G_{y}/\hbar v_{F}$. In the case of $\mathbf{M}$
directed along the $y$ axis (i.e., $G_{y}=G$), the eigen-energies of the
Hamiltonian are%
\begin{equation}
\varepsilon_{{b},\mathbf{p}}= d p^{2}  \pm \sqrt{(p_{x}-G)^{2}+p_{y}^{2}},
\label{e1}
\end{equation}%
where $\mathbf{p=}\hbar v_{F}\mathbf{k}$ is the momentum (in energy
representation) of the conduction ($b= c$) and valence ($b= v$) electrons
and $d = D/\hbar ^{2}v_{F}^{2}$. Note that the contact point energy corresponds to
$\varepsilon _{c,\mathbf{p}}=\varepsilon _{v,\mathbf{p}}= dG^{2}$ due to the
quadratic term. Accordingly, the relations $\varepsilon _{c,\mathbf{p} }
\geq dG^{2}$ and $\varepsilon _{v,\mathbf{p}} \leq dG^{2}$ are satisfied by
the conduction and valence bands, respectively.  The reference (i.e., zero) of energy is defied at the Dirac point of unmodified TI surface states.

Let us consider a normally incident electro-magnetic wave whose electrical
component $\mathbf{E}$ generates the electric dipolar transitions. The case
of monochromatic wave $\mathbf{E}=E_{0}\mathbf{\hat{e}}\exp {[i(\mathbf{q}%
\cdot \mathbf{r}-\omega t)]}$ corresponds to a certain polarization,
where $\omega $ is the angular frequency and $\mathbf{q}=q\mathbf{\hat{z}%
}$ is the photon wave vector. The circularly polarized wave is described by
polarization vector $\mathbf{\hat{e}=}$ $\mathbf{\hat{e}}_{\pm }=(\mathbf{%
\hat{x}}\pm i\mathbf{\hat{y}})/\sqrt{2}$, while $\mathbf{\hat{e}}_{\theta }=(%
\mathbf{\hat{x}}\cos \theta +\mathbf{\hat{y}}\sin \theta )$ corresponds to
linear polarization with the angle $\theta $ between the polarization plane
and the $x$ axis. The effect of unpolarized light can be described as a half
sum of the contributions induced by non-coherent waves with perpendicular
linear polarizations or opposite circular polarizations.

In the Coulomb gauge, the vector potential for the electro-magnetic
radiation has the form $\mathbf{A}=$Re$(ic\mathbf{E}/\omega )$, where $c$
is the velocity of light. From the perturbation treatment, the Hamiltonian of electron interaction with the radiation is
\begin{equation}
V=\frac{ev_{F}}{c}[\vec{\mathbf{\sigma }}\times \mathbf{A}]\cdot \mathbf{%
\hat{z}}+\frac{2eD}{\hbar c}\mathbf{k}\cdot \mathbf{A}\,,  \label{V1}
\end{equation}%
where $e$ is the electron charge. Neglecting the small term
containing $D$, the squared absolute value of electron-photon matrix
element is given as $|M|^{2} = v_{F}^{2}e^{2}E_{0}^{2}/8 \omega^{2}$
for both $\mathbf{\hat{e}}_{+}$ and $\mathbf{ \hat{e}}_{-}$
polarization.~\cite{Nair2008}    Thus, this is also
applicable to unpolarized light. The effect of linear polarization
appears in $|M|^{2}$ as the additional factor $2\sin ^{2}(\varphi _{p}-\theta )$ ($ \varphi _{p}$ is the angle between $\mathbf{p}$ and $\mathbf{\hat{x}}$),
which reduces to 1 after averaging over $\theta $. An important property of
the matrix element $M$ is its independence on the electron momentum and
exchange interaction.

The rate of photon energy absorption $P=\hbar \omega \sum_{\mathbf{p}}(W_{v,\mathbf{p};c,\mathbf{p}} - W_{c,\mathbf{p}; v,\mathbf{p}})$ is associated with the probability of interband transition $\left\vert b,\mathbf{p}\right\rangle \rightarrow \left\vert b^{\prime },
\mathbf{p}\right\rangle $:
\begin{equation}
W_{b,\mathbf{p};b^{\prime },\mathbf{p}}=\frac{2\pi }{\hbar }%
|M|^{2}f_{0}(\varepsilon _{b,\mathbf{p}},\mu )[1-f_{0}(\varepsilon _{b^{\prime},%
\mathbf{p}},\mu )]\delta (\left\vert \varepsilon _{b^{\prime },\mathbf{p}%
}-\varepsilon _{b,\mathbf{p}}\right\vert -\hbar \omega ).  \label{e3}
\end{equation}%
Here we ignore the small photon momentum $\mathbf{q}$ compared to $\mathbf{k}
$ as well as a minor deviation of the population factor $f({v,\mathbf{p}})$ from the equilibrium Fermi-Dirac function $f_{0}(\varepsilon
_{v,\mathbf{p}},\mu )$. Applying Eq.~(\ref{e1}) to Eq.~(\ref{e3}),
straightforward integration of $W_{v,\mathbf{p};c,\mathbf{p}}-W_{c,\mathbf{p}%
;v,\mathbf{p}}$ over $\mathbf{p}$ expresses the absorption rate $P$ in terms
of the roots of the energy $\delta $ function $p_{x}=p_{\lambda }\equiv G \pm \sqrt{(\hbar \omega /2)^{2}-p_{y}^{2}}$ ($\lambda =\pm $) as well as the residual integral over $p_{y}$:
\begin{equation}
P(\hbar \omega,\mu )=\frac{\alpha }{8}P_{i}\int\limits_{-\hbar \omega
/2}^{\hbar \omega /2}\sum\limits_{\lambda }\frac{f_{0}\left(
\varepsilon _{v,\mathbf{p}_{\lambda }},\mu \right) -f_{0}\left( \varepsilon
_{c,\mathbf{p}_{\lambda }},\mu \right) }{\sqrt{(\hbar \omega
/2)^{2}-p_{y}^{2}}}dp_{y}\,,  \label{3a}
\end{equation}%
where $\alpha =e^{2}/\hbar c$ is the fine-structure constant and $ P_{i} = (c/8\pi) E_{0}^{2}$ is the energy flow of the incident light. This
equation allows simplification under the assumption of relatively small energy shift $dG^{2}$ compared to the thermal energy $k_{B}T $. The final output in the normalized form (i.e., {\em normalized absorption}) can be written as
\begin{equation}
\frac{P(\hbar \omega,\mu )}{P_{i}}=\frac{\pi \alpha }{4}F(\nu ,\eta);
\;F(\nu ,\eta )=\frac{\sinh \nu }{\cosh \eta +\cosh \nu },  \label{3b}
\end{equation}%
in terms of arguments $\nu =\hbar \omega /2k_{B}T$ and $\eta =(4\mu
-d\hbar ^{2}\omega ^{2})/4k_{B}T$. The factor $\frac{1}{4}\pi \alpha
F(\nu ,\eta )$ determines the absorption ratio. At low temperature
($ \nu >>1$) in the absence of band curvature ($d=0$), this value is
four times smaller than the graphene absorbance $\pi \alpha $
$\approx ~2.3\%$,~\cite{Nair2008}   reflecting the absence of
degeneracy on valleys and spins. It is also obvious that $\pi \alpha
/4$ is the maximal structure absorbance attainable with a small
$\left\vert \mu \right\vert $ and high photon
energy.~\cite{XZhang2010}

Extending the analysis to the PGE begins with the general expression for the
surface current in the diffusive regime
\begin{equation}
j=-\frac{e}{A_{0}}\sum_{b,\mathbf{p}}f(b,\mathbf{p})v_{x}(b,\mathbf{p}),
\label{1}
\end{equation}%
where the axis $x$ is directed along the electron flow, $A_{0}$ is the area
of the TI/FM interface, and $\mathbf{v}(b,\mathbf{p})=v_{F}\nabla _{\mathbf{%
p}}\varepsilon _{b,\mathbf{p}}$ is the electron group velocity. The electron
distribution function $f(b,\mathbf{p})$  contains the non-equilibrium
effect due to light absorption [i.e., Eq.~(\ref{e3})].  When $f(b,\mathbf{p}) = f_{0}(\varepsilon_{b,\mathbf{p}},\mu)$ (i.e., in equilibrium), the current $j$ becomes zero exactly.  This also means the absence of dark current independent of the energy band distortions.  Accordingly, we hereinafter focus on the deviation from the equilibrium distribution; i.e.,  $\Delta f(b,\mathbf{p}) =f(b,\mathbf{p}) -f_{0}(\varepsilon_{b,\mathbf{p}},\mu)$.


In the stationary case, the relaxation time approximation  can yield
a simple solution to the kinetic equation for the distribution
function as  $\Delta f(b,\mathbf{p})=S(b,\mathbf{p})\tau
(b,\mathbf{p})$, where $\tau (b,\mathbf{p})$ is the relaxation time
and $S(b,\mathbf{p})$ is the rate of electron generation in the
state $|b, \mathbf{p}\rangle$. In turn, the generation rate can be
expressed via the probability of transitions discussed earlier [see
Eq.~(\ref{e3})]; i.e., $S(b,\mathbf{p})=\sum_{b^{\prime }}\left(
W_{b^{\prime },\mathbf{p};b,\mathbf{p}}-W_{b,\mathbf{p;}b^{\prime},
\mathbf{p}}\right) $. To extract the explicit effect of excitation
imbalance in the electrons with opposite group velocities, it is
sufficiently to restrict the consideration to a constant relaxation
time $\tau (b,\mathbf{p)} =\tau $ independent of momentum
$\mathbf{p}$\ and energy band $b$.~\cite{Hosur2011}    As
such, the population imbalance is solely defined by photo-transition
rates $S(b, \mathbf{p})$. At a fixed $p_{y}$ and the photon energy
$\hbar \omega $ ($ >2 |p_{y}|)$, the only states with $p_{x}=p_{\pm
} $ are involved in the interband transitions, while the proximity
effect produces the energy difference $\varepsilon
_{b,\mathbf{p}_{+}}-\varepsilon_{b,\mathbf{p}_{-}} = 4dG\sqrt{(\hbar
\omega /2)^{2} -p_{y}^{2}}$ for both electron and valence bands
(Fig.~1). This asymmetry between $  \varepsilon _{b,\mathbf{p}_{+}}$
and $\varepsilon _{b,\mathbf{p}_{-}}$ presents the main reason for
direct photocurrent generation. Indeed, it can be shown after some
algebra that the summation over $p_{x}$  in Eq.~(\ref{1}) produces
an expression determined by the difference of the population factor
rather than the velocities of photo-generated electrons and holes.
The net photocurrent effect gives the integral over $ p_y $ for a
given $\hbar \omega $ provided that $| \Delta f(b,\mathbf{p}) | \ll
f_{0}(\varepsilon_{b,\mathbf{p}},\mu)$:
\begin{equation}
j=\frac{\tau e}{2\pi \hbar ^{3}v_{F}}|M|^{2}  \int\limits_{-\hbar \omega /2}^{\hbar \omega /2} dp_{y} \left\{ \left[ f_{0}\left( \varepsilon _{v,\mathbf{p}_{+} },\mu\right) - f_{0}\left( \varepsilon _{v,\mathbf{p}_{-} },\mu\right) \right] - \left[ f_{0}\left( \varepsilon _{c,\mathbf{p}_{+} },\mu\right) - f_{0}\left( \varepsilon _{c,\mathbf{p}_{-} },\mu\right) \right] \right\}.  \label{4}
\end{equation}

Along with numerical evaluation, an analytical approximation of $j$ would be desirable for a qualitative analysis of the effect.  Assuming that the temperature is not too low ($k_{B}T > dG^{2}$), Eq.~(\ref{4}) can be represented in terms of the incident light power and function $F$:
\begin{equation}
j=-\frac{\pi \alpha }{4}dG\frac{ev_{F}\tau }{k_{B}T}P_{i}F(\nu ,\eta )F(\eta
,\nu ),  \label{5}
\end{equation}%
which explicitly displays the dependence of the PGE  on
absorption [Eq.~(\ref{3b})].  This expression clearly illustrates that
the photocurrent originates from both the proximity effect ($\sim G$),
which breaks the isotropy of the interface Hamiltonian, and the spin
independent parabolicity ($\sim d$) of the energy bands.
Furthermore, the dependence on photon energy (i.e., dimensionless
variable $\nu $) manifests a well-defined maximum that can be tuned
by a change in the chemical potential. Note that switching in the FM magnetization  ($ \mathbf{G} \rightarrow - \mathbf{G}$) leads to reversal
of the photocurrent, which is in accordance with time-reversal
symmetry inherent in the linear PGE.~\cite{Ivchenko95}

The photocurrent calculated in the structure with a typical TI is shown in Fig.~2 for different excitation energies at 300~K and 4~K. Specifically, Bi$_{2}$Se$_{3}$ is assumed along with $\tau = 1$ ps, $G=40$ meV, and the  incident radiation power density of $P_{i}=1$ W/cm$^{2}$.  The maximum current density at room temperature is of the order of $0.1~\mu$A/cm, whereas it reaches over 10~$\mu$A/cm at liquid helium temperature.  These numbers are drastically larger than the photocurrent reported earlier in
TIs.~\cite{Hosur2011,Mclver2011}   Another remarkable
property of the proposed PGE is its correlation with the normalized
absorption curves. As shown in Figs.~2(a) and 2(b),  stronger and broader
absorption characteristics leads to a similar trend in the
photocurrent response at room temperature.  On the other hand,  the
sharp transitions in the absorption curves (vs.\ $\mu$) at low
temperatures lead to the corresponding resonant current peaks [see
Figs.~2(c) and 2(d)]. This can be understood clearly from Figs.~1(b) and 1(c); note also the relation $\partial P /\partial\eta \propto j$ from
Eqs.~(\ref{3b}) and (\ref{5}) although its validity is limited at
low temperatures.

As indicated earlier, the main contribution to the photocurrent
comes from the asymmetrical distribution of the photo-excited
carriers that is in turn due to the shift $G$ of the Dirac cones and
the $Dk^{2}$ term in the Hamiltonian. While the parabolicity is
proportional to the parameter $D$, which is well-known for most TIs,
the magnitude of the exchange energy may vary depending on the TI/FM
materials as well as the quality of the interface. Based on the
earlier studies in the
literature,~\cite{Tse10,Semenov2008,Haugen2008,Chakhalian2006}
it appears that $ G= 5-50 $~meV can be expected.  In the limit of
low $G$, Eq.~(\ref{5}) predicts proportionality of the PGE to $G$.
However, numerical evaluation of the more general expression
[Eq.~(\ref{4})] illustrates a saturating trend in the peak
magnitudes of resonant curves $j(\mu )$  as $G$ increases ($dG^2 \gg
k_B T$). This is clearly visible in Fig.~3(a), where increasing $G$
from 5 meV to 40 meV at $T= 4$ K significantly broadens the resonant
photo-response but introduces only a minor change in the peak
height.  The saturating behavior can be explained, at
least in part, by the corresponding photon absorption curves.  As
can be seen from Fig.~3(b), an increase in $G$ leads to a more
gradual change in $P$ with a decreased slope. Since both factors
(i.e., $G$ and $\partial P /\partial \eta$) contribute, their impacts
compensate each other and the peak current remains largely
unaffected.


Figure~3(c) shows the position (in $\mu$) of the peak current
as a function of the photon energy.  The observed dependence is nearly
linear in a wide range of photon energy, suggesting the equivalence
(or strong correlation) between the frequency scanning and the
chemical potential (thus, gate bias) sweeping in the detection of
the PGE. In other words, the energy of incoming photons can be
determined by measuring the TI surface current as a function of the
gate bias.  The strength of the photo-response is particularly
pronounced at low temperatures as shown in Fig.~3(d). Note that all
curves are calculated assuming an identical relaxation time $\tau =1$
ps.  As the relaxation rate tends to decrease rapidly with
temperature, further enhancement of PGE beyond tens of $ \mu
$A$\cdot $cm/W appears likely at $T=4$ K.  Moreover, the relaxation
time can be much longer than 1 ps even at room temperature as the
sample quality improves.

The strong PGE discussed above can be used to realize a
photo-detector in the THz/far-infrared frequencies. One clear
advantage of this mechanism is that the dark current can be
eliminated without sacrificing the response time since the resulting
photocurrent does not require an external bias. In conventional
photo-detectors (such as those based on p-n junction photodiodes,
Schottky barriers, quantum well structures, etc.), an external bias is
applied to accelerate the excited carriers or to decrease the pn
junction capacitance for fast response. However, the consequential dark current is detrimental to the device performance at room temperature as it is a powerful source of noise. The prospective device, on the other hand, is
expected to work at room temperature with an extraordinary
sensitivity, while having a very short response time. Since no
junction is formed in the channel, the response time is determined
by the carrier transit time.  Not only the excited carriers have
a high initial ensemble velocity ($\sim v_{F}$) similar to the Schottky-barrier detector, but also the momentum relaxation in ideal TI surface states is expected to be very long, particularly for low energy electrons, due to the
suppression of backscattering. Overall, the structure offers a unique opportunity for room-temperature detection of long-wavelength
photons.

In summary, a linear PGE on the topological insulator (TI) surface
adjacent to an insulating FM is examined. The dc photo-excitation
stems from the TI surface band modification under the influence of
both the symmetry-breaking proximate exchange interaction and
spin-independent parabolicity of the dispersion law.  The phenomenon
is invariant to the sign of light circular polarization, thus
effective at unpolarized excitation. At the same time, the PGE
reveals the frequency-dependent resonance-like features,
particularly for low temperatures, despite the absence of any
discrete energy levels. The strong correlation between the
excitation energy and photo-response enables straightforward
detection of radiation frequency via electrical means such as the
gate bias.~\cite{Chen2010}   The photocurrent estimated with a
relatively short relaxation time of 1 ps exceeds those of the
previously reported PGEs by orders of magnitude.  The proposed TI/FM
structures may have significant advantages over the conventional
devices in the detection of long-wavelength photons beyond the thermal
noise limit.~\cite{Apell2012,Thongrattanasiri2012}

This work was supported, in part, by the SRC Focus Center on Functional
Engineered Nano Architectonics (FENA) and the US Army Research Office.

\newpage

\begin{center}
\begin{figure}[tbp]
\includegraphics[scale=.8,angle=0]{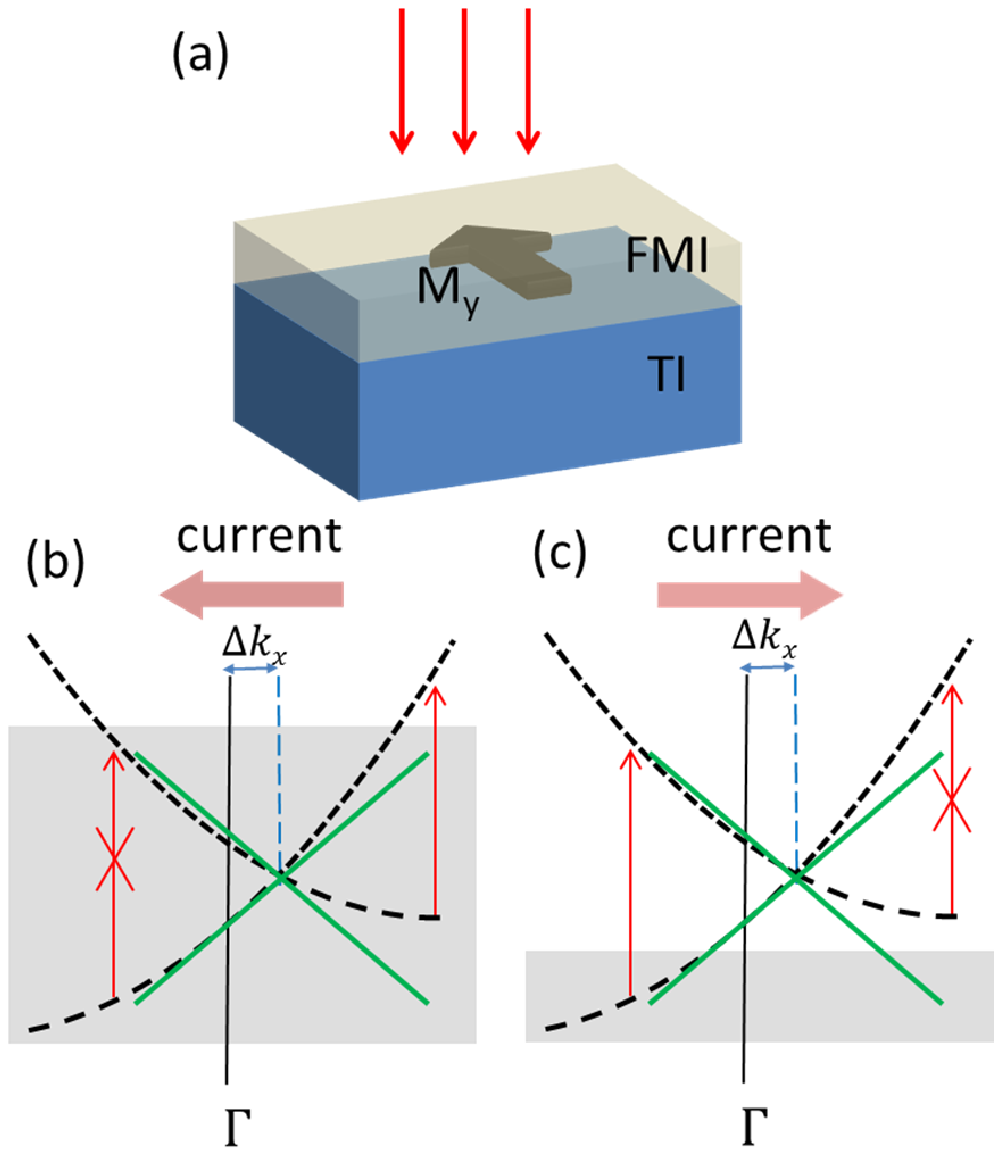}
\caption{(Color online) Schematic illustration of photocurrent generation
on a TI surface adjacent to an insulating ferromagnet (FM). (a) TI/FM hybrid
device structure.  The arrows indicate the incident electro-magnetic
radiation and the block arrow denotes the magnetization of the FM layer. (b,c)
Photocurrent generation in a n-type TI and a p-type TI, respectively.
The dashed lines are the TI surface bands taking into account the $k^2$
term and the proximity effect, while the solid lines only include the
proximity effect.  The shaded area denotes the filled states under
electrochemical potential $\mu$.}
\end{figure}
\end{center}

\newpage

\begin{center}
\begin{figure}[tbp]
\includegraphics[scale=.7,angle=0]{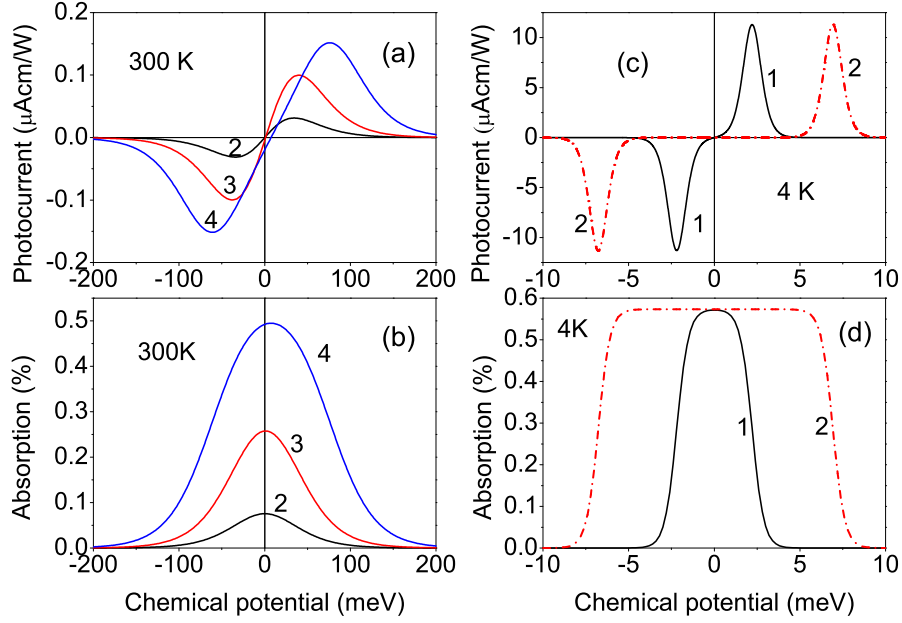}
\caption{(Color online) Photocurrent density and normalized
absorption curves at (a,b) 300~K and (c,d) 4~K versus the chemical
potential with different excitation energies:~\cite{Diehl09}  $\hbar \protect\omega = 4.4$ meV, 13.7 meV,
50 meV, and 135 meV for curves 1, 2, 3 and 4, respectively. All
calculations assume the relaxation time $\tau$ of 1 ps and the
proximate exchange energy $G$ of 40 meV along with the material
parameters corresponding to Bi$_{2}$Se$_{3}$ ($v_{F}=4.28\times
10^{7} $ cm/s and $D=13$~eV$\cdot\mathring{A}^{2} $).}
\end{figure}
\end{center}

\newpage

\begin{center}
\begin{figure}[tbp]
\includegraphics[scale=.95,angle=0]{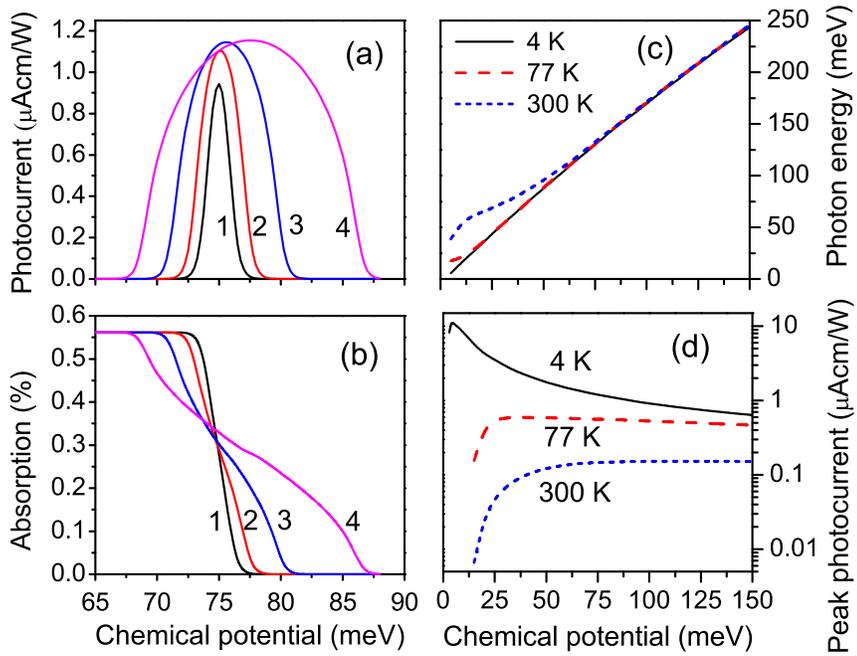}
\caption{(Color online) (a) Resonance-like photocurrent and (b) normalized absorption curves at 4~K with the excitation energy of 135 meV.  Curves 1, 2, 3, and 4 represent the results with the proximate exchange energy $G$ of 5 meV, 10 meV, 20 meV, and 40 mev, respectively. (c) Photon energy vs.\ peak position in the chemical potential $\mu$ and (d) corresponding peak photocurrent at different temperatures. $G = 40$ meV is used for (c) and (d).}
\end{figure}
\end{center}

\end{document}